\documentclass[a4paper,12pt,aps,prb,showpacs,showkeywords]{revtex4}
\usepackage{amsmath}
\usepackage{graphicx}

\begin{document}

\title{Electronic and hole minibands in quantum wire arrays  of different crystallographic structure }
\author{M. Krawczyk$^{1}$, J. W. Klos$^{1,2}$}
\address{$^{1}$Surface Physics Division, Faculty of Physics, Adam
Mickiewicz University,  Umultowska 85, 61-614 Poznan, Poland\\
$^{2}$Department of Science and Technology, Linköping University
601 74, Norrköping, Sweden}

\vspace{0.2cm}

\begin{abstract}
We consider quantum wire arrays consisting of GaAs rods embedded in
Al$_{x}$Ga$_{1-x}$As and disposed in sites of a square or triangular lattice. The
electronic and hole spectra around the conduction band bottom and the valence band
top are examined versus geometry of the lattice formed by the rods, concentration of
Al in the matrix material, and structural parameters including the filling
fraction and the lattice constant. Our calculations use the envelope function and
are based on the effective-mass approximation. We show that the electronic and hole
spectra resulting from the periodicity of the heterostructure, depend on the factors
considered and that the effect of lattice geometry varies substantially with lattice
constant. For low lattice constant values the minigaps are significantly wider in the case of
triangular lattice, while for high lattice constant values wider minigaps occur in
the square lattice-based arrays. We analyse the consequences of our findings for the
efficiency of solar cells based on quantum wire arrays.

\end{abstract}
\maketitle

\section{Introduction}

Semiconductor periodic heterostructures have an additional, superstructure-related
discrete translational symmetry, the period of which is much larger than that of the
atomic lattice. This causes electronic and hole minibands, with minigaps between
them, to form near the bottom of the conduction band and the top of the valence
band, respectively, in the electronic/hole spectrum of the structure. In
superlattices, heterostructures periodic in a one dimension (1D), the
electronic/hole band splits at the boundary and in the centre of the first Brillouin
zone to form minibands separated by minigaps of width equal to the difference
between the conduction band bottom or the valence band top, respectively, in the
component materials of the superlattice.\cite{davison92} Similar effects occur also
in quantum wire arrays, semiconductor heterostructures periodic in two dimensions
(2D), but the opening of minigaps and their width, determined in a more complex way
by factors which include structural parameters in addition to material parameters,
can only be studied numerically. The effects in question are also observed in other
periodic composites, such as photonic, phononic or magnonic crystals, in the
dispersion of electromagnetic, elastic or spin
waves.\cite{yablonovitch87,john87,siglas93,vasseur94,vasseur08,vasseur96,krawczyk08,tkachenko09}

Aroused many years ago and persisting ever since,  the interest in periodic
semiconductor heterostructures stems from the possibility of fabricating systems
with the conduction or valence band split into a desired pattern of discrete
minibands and minigaps\cite{bryant89,nika07,gershoni93};
this could be achieved through systems of semiconductor dots or rods of properly designed
size and shape (cylinders, cubes, prisms, rings or antirings)\cite{cusack96,tkach02,planelles06,planelles07}. The discrete spectra of
electronic conduction or valence bands significantly modify the electron or hole
transport as well as the optical properties of the structure. Used in photovoltaic
and thermoelectric cells, semiconductor heterostructures allow construction of
devices converting light\cite{barnham97,green03,nozik02,jiang06,shao07} or
temperature difference\cite{balandin03,lin03,yang04,broido06} into electric
current with efficiency higher than that allowed by homogeneous materials. Although
the two effects differ in nature, the electronic or hole band structure plays an
important role in both. The numerous theoretical studies on the subject use various
approximations and methods\cite{allmen92,harrison05,mingo07,gomez02,lazrenkova01,broido06,duque99} of
determining the energy spectrum of 2D and 3D heterostructures. Here we use the plane wave method with the
effective-mass approximation, a technique successfully applied to studying the
electronic states in 1D, 2D and 3D heterostructures with quantum dots and wires of
different shape and size, as well as interdiffusion and strain effects on electronic
bands.\cite{cusack96,gershoni88,li98,li05,tkach00,ngo06}

The main objective of this study is to investigate the effect of the
crystallographic structure on the electronic/hole spectrum in quantum wire arrays.
In particular, we are going to determine the conditions conducive to the opening of
minigaps in the conduction band and in the valence band. Besides its purely
scientific aspect, our investigation is of practical importance for photovoltaic
cells, the efficiency of which is determined by the electronic structure of the
active material in which the photoelectric effect takes place. If the active
material is a periodic semiconductor heterostructure, the efficiency of the cell
will substantially depend on the width of the electronic energy minigap between the
lowest miniband and the rest of the conduction band.\cite{klos09} 

The paper is organized as follows. In Section \ref{pwm} we present in detail the
plane wave method applied to the equation for the electronic envelope function in
quantum wire arrays, and discuss the approximations used. Section \ref{C_B}
discusses the results of our calculations obtained for the conduction band in a
quantum wire array consisting of GaAs rods embedded in AlGaAs and forming a square
or triangular lattice. The spectra of these 2D heterostructures  are studied versus
Al concentration in the matrix material, filling fraction and lattice constant. In
Section \ref{V_B} we analyse the energy spectrum of the valence band in the same
heterostructures. The paper is summed up in Conclusions, in which we indicate the
implications of our results for the efficiency of solar cells based on quantum wire
arrays.

\section{Plane wave method in effective-mass approximation} \label{pwm}

Let us consider a quantum wire arrays consisting of infinitely long rods of material
A forming a square or triangular lattice and embedded in a matrix of material B
(Fig. 1). The rod axes are oriented along the $z$ direction. The rods are assumed to
represent potential wells for propagating electrons and holes.


Here we shall only consider arrays with rods of circular cross section. The effect
of the cross-sectional geometry of the rods has already been studied in our earlier
paper \cite{klos09}, in which we established that the efficiency of solar energy
conversion in the photoelectric effect only slightly depends on the shape of the
rods and that their circular cross section is optimum in this aspect. The filling
fraction $f$, defined as the area ratio of the rod cross section to the unit cell of
the heterostructure, will be a measure of the rod size. The maximum filling fraction
value corresponds to the situation in which rods touch the unit cell limits. For a
square or triangular lattice of cylindrical rods with cross-section radius $\rho$
the filling fraction can be calculated from the following respective formulae:
\begin{equation}
f= \frac{\pi \rho^{2}}{ a^{2}} \text{  and  }  f = \frac{ \pi \rho^{2}}{\sqrt{3} a^{2}}, \label{filling_fraction}
\end{equation}
where $a$ is the distance between adjacent lattice sites.

Our calculations of the  electronic band structure of 2D semiconductor
heterostructures will be performed in the effective-mass approximation\cite{bustart88,burt99,califano00}, which means that we shall restrict our
attention to the bottom (and top) of the parabolic conduction (and parabolic
valence) band in the electronic (hole) spectrum of the constituent semiconductors.
Such approximations prove to be very useful in the electronic band calculations for
semiconductor heterostructures with various component materials, as well as for
single dots.\cite{califano00,harrison05,tablero05,califano00b,duque99}
We shall assume the interactions between the single conduction band and the two
valence bands are negligible; this allows independent determination of the
electronic miniband structure in the conduction band (Sections \ref{pwm} and
\ref{C_B} in the single band approximation) and the hole miniband structure in the
valence band (Section \ref{V_B}).

Electronic states are described by the Ben-Daniel-Duke equation:\cite{duke}
\begin{eqnarray}
\left[-\alpha\left(\nabla \frac{1}{m^{*}(\mathbf{r})}\nabla
\right)+E_{C}(\mathbf{r})\right]\Psi_{e}(\mathbf{r})=E\Psi_{e}(\mathbf{r}),
\label{eq:r1}
\end{eqnarray}
with effective mass $m^{*}(\mathbf{r})$ (isotropic in a homogeneous medium) and
effective potential $E_{C}(\mathbf{r})$ determining the position of the conduction
band bottom; ${\bf r}$ denote the position vector in the plane of the periodicity;
$\Psi_{e}$ denotes the envelope wave function of conduction band electrons. The
constant $\alpha =10^{-20}\hbar^{2}/(2 m_{e} e)\approx 3.80998$ (where $m_{e}$ and
$e$ denote the free electron mass and charge, respectively) allows to express the
energy $E$ in electronvolts (eV) and the coordinates $x$, $y$ in angstroms {\AA}.
The material parameters: the effective mass and the effective potential, vary in
space with the periodicity of the heterostructure:
\begin{eqnarray}
m^{*}({\bf r}+{\bf R})&=&m^{*}({\bf r}),\nonumber\\
E_{C}({\bf r}+{\bf R})&=&E_{C}({\bf r}),\label{eq:r2}
\end{eqnarray}
where ${\bf R}$ is a lattice vector (${\bf R} = a(m_{x},m_{y})$ in square and ${\bf
R} = a(m_{x}+m_{y}/2,\sqrt{3} m_{y}/2)$ in triangular lattice, $m_{x}$ and $m_{y}$
being integers).

The expansion of the electron envelope function in the plane-wave basis and the
Fourier expansion of the material parameters can be written as:
\begin{eqnarray}
\Psi_{e}({\bf r})&=&\sum_{{\bf G}}\phi_{e}^{\bf G}e^{i ({\bf G}+{\bf k})\cdot{\bf r}},\nonumber\\
E_{C}({\bf r})&=&\sum_{{\bf G}}E_{C}^{\bf G}e^{i {\bf G}\cdot{\bf r}},\nonumber\\
w({\bf r}) \equiv 1/m^{*}({\bf r})&=&\sum_{{\bf G}}w^{\bf G}e^{i {\bf G}\cdot{\bf
r}},\label{eq:r7}
\end{eqnarray}
where $\phi_{e}^{\bf G}$ are Fourier coefficients of the periodic factor of the
envelope function (which has the same periodicity as the quantum wire arrays) and
$\bf k$ is a wave vector from the first Brillouin zone. Fourier coefficients of the
material parameters, $E_{C}^{\bf G}$ and $w^{\bf G}$, can be found analytically from
the formula:
\begin{eqnarray}
F^{\bf G}=\frac{1}{V}\int_{V} f({\bf r})e^{-i{\bf G}\cdot{\bf r}}d{\bf
r},\label{eq:r8}
\end{eqnarray}
where $f({\bf r})$ and $F^{\bf G}$ denote, respectively, the periodic material
parameter  (the effective mass or the effective potential) and the corresponding
Fourier coefficient for a plane wave of wave vector equal to vector ${\bf G}$ of the
reciprocal lattice (${\bf G}= (2 \pi / a) (n_{x},n_{y})$ for square lattice, ${\bf
G}= (2 \pi / a) (n_{x},(2 n_{y}-n_{x})/ \sqrt{3})$ for triangular lattice, $n_{x}$
and $n_{y}$ being integers); $V$ denotes the area of the periodic heterostructure
unit cell. The explicit form of the Fourier coefficients (\ref{eq:r8}) for rods of
circular cross section is:
\begin{eqnarray}
\label{eq10} F^{\bf G} = \left\{ {\begin{array}{l}
 F_{A} f - F_{B} (1 - f) \;\;\; {\rm for }\;\; {\bf G} = 0, \\
 (F_{A} - F_{ B} )2f \frac{J_1 (G \rho)}{G \rho}\;\; {\rm for }\;\; {\bf G} \ne 0,
\end{array}} \right. \label{r:F}
\end{eqnarray}
where functions $F_{A}$ and $F_{B}$ are the values of potential  ($E_{C,A}$ and
$E_{C,B}$) or effective mass inverse ($1/m^{*}_{A}$ and $1/m^{*}_{B}$) in the
cylinders and in the matrix, respectively, and $J_{1}$ is a Bessel function of the
first kind.

The substitution of expansions (\ref{eq:r7}) in the Schr\"{o}dinger equation
(\ref{eq:r1}) leads to the following system of equations representing an eigenvalue problem,
the solution of which yields Fourier coefficients of the periodic factor of the
envelope function and the electron energy $E$:\cite{dobrzynski95,cusack96}
\begin{eqnarray}
\sum_{{\bf G'}}\Big[\alpha\left({\bf G}+{\bf k}\right)\cdot\left({\bf G'}+{\bf
k}\right)w^{{\bf G'}-{\bf G}} +E_{C}^{{\bf G'}-{\bf G}}\Big]\phi_{e}^{{\bf
G'}}=E\phi_{e}^{{\bf G}}.\label{eq:r9}
\end{eqnarray}
When a finite number $N$ of reciprocal lattice vectors is used in the Fourier series
(\ref{eq:r7}) also the system of equations (\ref{eq:r9}) becomes finite.  We have
solved this system of equations by standard numerical procedures designed for
solving symmetrical matrix eigenvalue problems, and tested the eigenvalues found for
convergence, as necessary in procedures of this type. For all the structures
considered in this study a satisfactory convergence of numerical solutions of Eq.
(\ref{eq:r7}) for first few bands proves to be attained with the use of 169 reciprocal lattice vectors.

\section{The effect of the crystallographic structure on the conduction band}\label{C_B}

The results presented in this Section have been obtained for rods of  gallium
arsenide (GaAs) embedded in a matrix of gallium arsenide doped with aluminium
(Al$_{x}$Ga$_{1-x}$As, where $x$ is the concentration of Al ions). In this case the
depth of the potential wells is determined by the concentration of Al in the matrix:
growing matrix Al concentration will increase the effective potential in the matrix,
and thus deepen the potential wells felt by conduction band electrons. We have used
the following empirical formulae for a linear extrapolation of the material
parameter values in GaAs and AlAs to estimate their values in the
Al$_{x}$Ga$_{1-x}$As matrix: $E_{C} = 0.944 x$ and $m^{*} = 0.067 + 0.083
x$.\cite{param,param2}

The effective-mass approximation is justified by the occurrence of a direct gap at
point $\Gamma$ of the atomic band structure for low concentrations of Al in both the
rods and the matrix. To meet this condition of approximation applicability, the Al
concentration in the matrix was only allowed to range from 0 to 0.35. Also
calculated in the envelope function approximation and by the plane wave method, the
electronic band structure (the conduction band) of square lattice-based
heterostructures presented in [\onlinecite{bryant89,dobrzynski95,krawczyk09}] is in
agreement with the results obtained in this study from Eq. (\ref{eq:r9}).

Figure \ref{fig2x} (a) and (b) shows the energy structure of conduction band electrons versus Al concentration in a 2D heterostructure with GaAs rods disposed in sites of a triangular or square lattice. The calculations were performed at fixed lattice constant $a$ = 120 {\AA} and filling fraction $f=0.3$. Increasing $x$ results in deepening potential wells, in which electrons from first minibands localise and the interaction between electrons in neighborhood wells weakens. As a consequence the miniband shrinks and increase their energy, and the minigap that separates it from the quasi-continuous conduction band above grows wider. There are also some differences between lattices: in the triangular lattice case the first miniband is seen to occur almost throughout the considered range of Al concentration $x$ (from $x = 0.01$ up), while in the square lattice case the Al concentration must be above a critical value $x=0.07$ for the gap to open. Note from Al concentration $x=0.13$ up the first minigap is wider in the square lattice case than in the triangular lattice case. The second minigap (between the third and the fourth miniband) only opens at $x=0.22$ and $x=0.20$, for the square and triangular lattice, respectively, and is wider in the square lattice-based structure.

Let us see how the band structure varies with filling fraction. Figure \ref{fig:3f} shows plots obtained for matrix Al concentration fixed at $x = 0.35$ and for lattice constant $a = 120 $ {\AA} in the triangular and square lattice case. As expected, the minibands are seen to shift down as the filling fraction grows due to widening of the wells and to become wider because of shrinking of matrix (barrier) material\cite{krawczyk09}. The filling fraction dependence in both structures is very similar in terms of ranges of minigap occurrence as well as filling fraction values corresponding to the maximum width  of the first and second gap, $f \approx 0.18$ and $f \approx 0.39$, respectively (see Fig. \ref{fig:4}). Note, however, that in the case of rods forming a square lattice both the first minigap and the second one are wider, their respective widths being 0.17 eV and 0.1 eV, against 0.14 eV and 0.08 eV in the triangular lattice case. These small differences between electronic bands in square and triangular lattices in dependence on  $x$ and $f$ can be explained as follows.  


A major difference between a triangular lattice and a square  lattice is the number
of nearest neighbours; in a triangular lattice-based structure each well (rod) has 6
neighbours (at a distance $a$), against 4 neighbours (at a distance $a$) in the
square lattice case. A larger number of nearest neighbours implies more interactions, if exist,
and wider energy bands for the same parameter values, hence minibands  in the
triangular lattice case should be slightly wider than in a square lattice-based
structure with the same values of $x$, $f$ and $a$ (especially those with high energy, e. g. compare Figs \ref{fig2x} (a) and (b) or Figs \ref{fig:3f}(a) and (b)). Obviously, this affects the minigap width, in which differences are seen as
well, Fig. \ref{fig:4}. Let us get back to the definition of filling fraction,
(\ref{filling_fraction}), to determine the radii of the cylindrical rods
corresponding to equal filling fraction in both types of lattice at equal lattice
constant values. The cylindrical rods are found to have larger diameter in the
square lattice case ($\rho_{\text{square}} \approx 1.0746 \rho_{\text{triangular}}
$), which implies relatively lower energies of the lowest minibands associated with
electrons in wells, and consequently, increased isolation of these states, and
narrower bands, also conducive to wider minigaps. However, it should be kept in mind
that the spacing between rods decreases as their diameter grows, which can lead to
increased overlapping of the wave functions of adjacent rods, and consequently, to
wider minibands. The impact of these two competing effects can be expected to depend
on the lattice constant.


Let us compare plots depicting the band
structure versus the lattice constant in the square and triangular lattice case
(Fig. \ref{fig:5}). 
 In this dependence the electronic band
structures in the two cases considered differ the most. In the triangular
lattice case the first minigap is seen to occur throughout the considered lattice
constant range, i.e. from 25 {\AA} to 150 {\AA}, while in the square lattice case
the gap only opens above $a \approx 45$ {\AA}. For the second gap the critical
lattice constant values are $a = 87$ {\AA} and 92 {\AA} in the triangular and square
lattice case, respectively.
In Fig. \ref{fig:6} the width of minigaps in the square and
triangular lattice-based heterostructures is represented by the respective shapes,
i.e. squares and triangles. Filled and outlined symbols refer to the first and
second gap, respectively. For low lattice constant values a single gap is seen to
occur in the triangular lattice case; the width of this gap decreases monotonously
with increasing lattice constant from 0.3 eV for $a= 25$ {\AA} to 0.06 eV for $a =
250$ {\AA}. In the square lattice case the width of the first gap is seen to vary
with $a$ in a non-monotonous manner: from $a= 42.5$ {\AA} up the gap widens to reach
a maximum width of 0.16 eV at lattice constant 100 {\AA}, and then narrows down as
$a$ increases still further. For $a=75$ {\AA} the first gap has equal width in both
cases. The second gap has a maximum width for $a \approx 155$ {\AA} for both lattice
types considered, but tends to be wider in the square lattice case. Thus, for large
lattice constant values we observe the relations already explained above.

Let us try to elucidate the intriguing relations found in the range of low lattice constant
values. Two effects occur as a result of reducing the lattice constant at fixed
filling fraction: (a) the rod (potential well) diameter decreases, with a concurrent
increase energy of bands; at the same time, (b) the distance between adjacent wells
decreases, which implies increased overlapping of wave functions and thus steps up
the widening of bands. But it doesn't explain relations for $a < 55$ {\AA} where the top of the first miniband is already above the  barrier, for both arrangements and electron waves can propagate across the structure slightly scattered by the barrier. It means, that triangular lattice, due to a larger number of adjacent wells, favors destructive interference of such waves resulting in wide gap, as opposed to square lattice, where the gap disappear.    
Figure \ref{fig:7} shows the band structures,
calculated along the path $\Gamma=(0,0) \rightarrow \text{X}=2\pi/a (0.5,0) \rightarrow
\text{J}=2\pi /a(0.5,1/(2\sqrt{3}) ) \rightarrow \Gamma$ and $\Gamma=(0,0) \rightarrow \text{X}=2\pi/a (0.5,0) \rightarrow \text{M}=\pi/a (0.5,0.5) \rightarrow \Gamma$ in the first Brillouin zone, of
triangular (solid lines) and square (dashed line) lattice-based 2D heterostructure, respectively,  with lattice constant $a=50$ {\AA} (a) and
$a=120$ {\AA} (b). As expected the bands are seen to be very wide in the quantum wire arrays with
low lattice constant value, Fig.  \ref{fig:7} (a),
and very narrow in the arrays with large lattice constant Fig.  \ref{fig:7} (b). In Fig. \ref{fig:7} (a) we can observe additional difference in band structures for triangular and square lattices. For triangular lattice separating the first miniband form the second one at X point remains almost unaltered in comparison to point J. In the case of square lattice the separations both in X and M point are reduced as lattice constant decreases. This effect and some small shift of miniband position between points X and M cause the closing of absolute minigap in square lattice whereas for triangular lattice this minigap remains open.


\section{The effect of the crystallographic structure on the valence band}\label{V_B}

For calculating the valence band structure of quantum wire arrays we  use
the plane wave method, applied to the Schr\"{o}dinger equation of the envelope
function of light and heavy hole states near the top of the valence
band:\cite{datta08}
\begin{equation}
-\left( \begin{array}{cc}
\hat{Q}+\hat{P}&-\hat{R}^{*}\\
\hat{R}&\hat{Q}-\hat{P}
\end{array} \right)
\left( \begin{array}{c}
\Psi_{lh}({\bf r})\\
\Psi_{hh}({\bf r})
\end{array}
\right) =E\left( \begin{array}{c}
\Psi_{lh}({\bf r})\\
\Psi_{hh}({\bf r})
\end{array}
\right),\label{eq:r3}
\end{equation}
where $\Psi_{lh}({\bf r})$ and $\Psi_{hh}({\bf r})$ are envelope functions for light
and heavy holes, respectively, and operators $\hat{P}$, $\hat{Q}$ and $\hat{R}$ have
the form:
\begin{eqnarray}
P&=&E_{V}({\bf r})+\alpha \left( \frac{\partial}{\partial x}\gamma_{1}({\bf
r})\frac{\partial}{\partial x}+ \frac{\partial}{\partial y}\gamma_{1}({\bf
r})\frac{\partial}{\partial y} \right),
\nonumber\\
Q&=&\alpha \left( \frac{\partial}{\partial x}\gamma_{2}({\bf
r})\frac{\partial}{\partial x}+ \frac{\partial}{\partial y}\gamma_{2}({\bf
r})\frac{\partial}{\partial y} \right),
\nonumber\\
R&=&\alpha\sqrt{3}\left[ - \left( \frac{\partial}{\partial x}\gamma_{2}({\bf
r})\frac{\partial}{\partial x}- \frac{\partial}{\partial y}\gamma_{2}({\bf
r})\frac{\partial}{\partial y}
\right)\right.\nonumber\\
& & +\left. i \left( \frac{\partial}{\partial x}\gamma_{3}({\bf
r})\frac{\partial}{\partial y}+ \frac{\partial}{\partial y}\gamma_{3}({\bf
r})\frac{\partial}{\partial x} \right) \right].\label{eq:r4}
\end{eqnarray}
Luttinger parameters $\gamma_{1}$,  $\gamma_{2}$,  $\gamma_{3}$, describing the
effective masses $1/(\gamma_{1}+\gamma_{2})$ and $1/(\gamma_{1}-\gamma_{2})$ of
light and heavy holes, respectively, near point $\Gamma$ of the atomic lattice, are,
like the position of the valence band top $E_{V}$, periodic in the heterostructure.

Expanded in the plane-wave basis by a procedure similar to that employed in the case
of the conduction band, equation (\ref{eq:r3}) can be written as a Hermitian
eigenvalue problem for interacting light and heavy holes:
\begin{eqnarray}
\sum_{{\bf G'}}\Big[\left(A_{1}+A_{2}\right)\phi_{lh}^{{\bf G'}}+\left(B_{1}-iB_{2}\right)\phi_{hh}^{{\bf G'}}\Big]=E\phi_{lh}^{{\bf G}},\nonumber\\
\sum_{{\bf G'}}\Big[\left(B_{1}+iB_{2}\right)\phi_{lh}^{{\bf
G'}}+\left(A_{1}-A_{2}\right)\phi_{hh}^{{\bf G'}}\Big]=E\phi_{hh}^{{\bf
G}},\label{eq:r10}
\end{eqnarray}
where $A_{1}, A_{2}$ and $B_{1}, B_{2}$ are expressed as follows:
\begin{eqnarray}
A_{1}&=&-\alpha\gamma_{1}^{{\bf G}-{\bf G'}}\left({\bf G}+{\bf k}\right)\cdot\left({\bf G'}+{\bf k}\right)-E_{V}^{{\bf G}-{\bf G'}},\nonumber\\
A_{2}&=&-\alpha\gamma_{2}^{{\bf G}-{\bf G'}}\left({\bf G}+{\bf k}\right)\cdot\left({\bf G'}+{\bf k}\right),\nonumber\\
B_{1}&=&-\alpha\sqrt{3}\gamma_{2}^{{\bf G}-{\bf G'}}\left[(G_{x}+k_{x})(G'_{x}+k_{x})\right.\nonumber\\&&-\left.(G_{y}+k_{y})(G'_{y}+k_{y})\right],\nonumber\\
B_{2}&=&\alpha\sqrt{3}\gamma_{3}^{{\bf G}-{\bf
G'}}\left[(G'_{x}+k_{x})(G_{y}+k_{y})\right.\nonumber\\&&+\left.(G_{x}+k_{x})(G'_{y}+k_{y})\right],\label{eq:r11}
\end{eqnarray}
$\phi_{hh}^{\bf G}$ and $\phi_{lh}^{\bf G}$ have the sense of Fourier coefficients
for periodic factors of the envelope functions of heavy and light holes,
respectively, and $\gamma_{i}^{\bf G}$ ($i$ = 1, 2 or 3) and $E_{V}^{\bf G}$ are the
respective  Fourier coefficients of Luttinger parameters $\gamma_{i}$ and $E_{V}$,
which can be found from (\ref{r:F}) for the Luttinger parameter values, specified
below, in the rod material A and in the matrix material B.

Again, let us consider a heterostructure consisting of a triangular or square
lattice-based system of GaAs rods embedded in Al$_{x}$Ga$_{1-x}$As. The following
material parameter values, dependent on the concentration of Al in aluminium gallium
arsenide, can be assumed:\cite{param,param2}
\begin{eqnarray}
E_{V}&=&1.519+0.75 x,\nonumber\\
\gamma_{1}&=&6.85- 3.40 x,\nonumber\\
\gamma_{2}&=&2.10- 1.42 x,\nonumber\\
\gamma_{3}&=&2.90- 1.61 x.\label{eq:r6h}
\end{eqnarray}

Fig. \ref{fig:8} shows the wave vector dependence of hole minibands plotted
along the irreducible part of the first Brillouin zone in a triangular lattice-based
heterostructure (solid line, path $\Gamma \rightarrow \text{X} \rightarrow \text{J} \rightarrow
\Gamma$) and in a square lattice-based heterostructure (dashed line, path $\Gamma
\rightarrow \text{X} \rightarrow \text{M} \rightarrow \Gamma$). The calculations were performed
for lattice constant $a = 50$ {\AA}, filling fraction $f = 0.3$ and matrix Al
concentration $x= 0.35$.

Comparing the hole energy structure (Fig. \ref{fig:8}) with the  energy
structure of conduction band electrons (Fig. \ref{fig:7}), we notice the hole bands
(for light and heavy holes) are much narrower and arranged very densely, so their role in photoelectric effect will be minor. The first
three electronic bands lie in the energy range from 0.21 eV to 1.14 eV, against
-1.76 eV to -1.65 eV in the case of hole bands (the negative energy values are a
consequence of assuming the energy $E_{C,A}$ of a conduction band bottom electron as
the zero energy level). The lattice type-related differences in energy structure
will be better visualised in the dependence of hole minibands on the lattice
constant and the filling fraction.


The lattice constant dependence of the band structure in the triangular lattice case
is depicted in Fig. \ref{fig:9}(a). The minibands are seen to shift towards higher
energies and shrink as the lattice constant increases. A similar tendency is seen in
the square lattice case, in which, however, the minibands are shifted up: for
$a=250$ {\AA} the shift is 0.03 eV. Narrow bands are characteristic of weakly
interacting states localised in potential wells, and their large number is a
consequence of the low valence band maximum ($E_{V,B} = -1.78$ eV) in the matrix
material. The minigap width, plotted versus lattice constant in Fig.
\ref{fig:9}(b), is seen to grow steeply at first to reach a maximum and then
reduce monotonously as the lattice constant continues to increase. This behaviour is
typical of all the minigaps found in both the triangular lattice-based array and the
square lattice-based structure. Minor differences in minigap position and width only
occur for low lattice constant values and disappear as the lattice constant grows.


Let us scrutinize the evolution of the hole minibands with filling fraction. Figure
\ref{fig:10}(a) shows the highest two hole minibands versus filling fraction in the
triangular and square lattice case (lines with triangles and squares, respectively).
The bands are seen to shift towards higher energies with increasing $f$. In the
square lattice case this increase in energy is much stronger, which implies wider
minigaps, as depicted in Fig. \ref{fig:10}(b), showing the width of the two
minigaps plotted versus filling fraction. The minigaps are the widest ($\Delta E$ up
to 0.05 eV) for low filling fraction values, $f \approx 0.05$, and their maximum
width represents less than one third of that of electronic minigaps in the valence
band.


Increasing frequency of holes minibands with increase of lattice constant and
filling fraction is due to the same mechanism as has been already  explained in Sec.
\ref{C_B} for the frequency of electron minibands decrease. This is because
cylinders of GaAs forms wells for the holes in  valence band, as well as for
electrons in conduction band. The observed differences in both dependencies (for
electrons and holes) are mainly due to the number of states localized in these
wells. We can compare this numbers with the help of thick (red) lines in Figs
\ref{fig:7} and in Fig. \ref{fig:8} for electrons and holes
respectively. In the case of holes, the Al$_{0.35}$Ga$_{0.65}$As form the barier
with the height  -1.78 eV and the large number of hole minibands have energy  inside
the wells. This is responsible for similar behavior of hole minibands and minigaps
in the presented in Figs \ref{fig:9} and \ref{fig:10} dependencies on lattice
constant and filling fraction.

\section{Conclusions}

We have examined thoroughly the effect of the crystallographic structure on the
electronic and hole energy bands in quantum wire arrays by comparing structures
consisting of GaAs rods embedded in Al$_{x}$Ga$_{1-x}$As and disposed in sites of
square or triangular lattice. We have established that both the electronic and hole
minigaps tend to be wider in the square lattice-based structure in the considered
ranges of parameters that include the lattice constant, the filling fraction and the
potential well depth. Exceptions from this rule are heterostructures with low (less
than 75 {\AA}) lattice constant values (with the first electronic minigap, between
minibands 1 and 2, wider in the triangular lattice case than in the square lattice
case) and heterostructures with low Al concentrations in the Al$_{x}$Ga$_{1-x}$As
matrix. We have demonstrated that the conduction band bottom and the valence band
top can be controlled in the range $\sim$ 0.3 eV and $\sim$ 0.2 eV, respectively, by
adjusting the filling fraction, and in the range $\sim$ 0.14 eV and $\sim$ 0.1 eV,
respectively, by adjusting the lattice constant. Also the minigap width is shown to
be controllable, especially the width of the gap between minibands 1 and 2, which
can attain (at matrix Al concentration 0.35) 0.19 eV and 0.06 eV in the conduction
band and in the valence band, respectively.

The above-discussed properties of semiconductor heterostructures will be crucial for
the efficiency of solar cells based on these materials.
In monolithic semiconductors with a single bandgap the  absorption of photons is incomplete because photons of energy below the bandgap width are not absorbed by the system. On the other hand, thermalisation causes the surplus energy of electrons above the bottom of the conduction band or holes below the top of the valence band to dissipate in thermal contact with the crystal lattice. Photon absorption can be increased by using tandem cells - cascades of solar cells of successively narrowing bandgaps\cite{tandem,tandem2}. A similar effect can be obtained with solar-energy converters based on semiconductor heterostructures discussed in this paper  with multiple bandgaps, which also allow  to reduce  thermalisation losses.\cite{nozik02,jiang06,luq,34band,multiband,multiband2,multiband3}

AlGaAs heterostructures have relatively wide energetic gap (1.4-2.1eV) which make them potentially suitable for designing of solar cells with intermediate band, were the wider gap  (e.g. in comparison to the bulk Si) is required for achievement of high efficiency of solar radiation conversion. The key role in this gain in efficiency is played by the lowest conduction miniband, which, detached from the overlapping conduction minibands, acts as an intermediate band that opens an extra channel for carrier transitions between the valence band and the conduction band.  Another parameter of vital importance for the efficiency of solar-energy conversion is the distance between the top of the highest valence miniband and the bottom of the overlapping conduction minibands. This distance, determines the energy of utilized carriers.\cite{klos09} The effect of the other electronic minigaps on the
efficiency of solar-energy conversion remains to be investigated.  Considering that
for filling fraction $f > 0.5$ and/or sufficiently large lattice constant values
($a>160$ {\AA}  for $f=0.3$) the second minigap in the conduction band is comparable to or
even wider than the first one, it can be expected to affect the efficiency of
solar-energy conversion by providing additional allowed transitions. However, a
precise evaluation of the effect of further discretisation of the conduction or
valence band is beyond the scope of this study and will be the object of our future
research. The impact of the discretisation of the valence band on the efficiency of the
photovoltaic effect will be much lesser than that of the discretisation of the
conduction band, and, in the first approximation, can be regarded as limited to a
shift in valence band top in the evaluation of solar cell efficiency.\cite{klos09}

\subsection*{Acknowledgements}
This study was supported by grant No. N~N507~3318~33 of the Polish Ministry of
Science and Higher Education. The calculations presented in this paper were
performed in  Pozna\'n Supercomputing and Networking Center.

\begin{figure}
\centering
\includegraphics[width=3.5in]{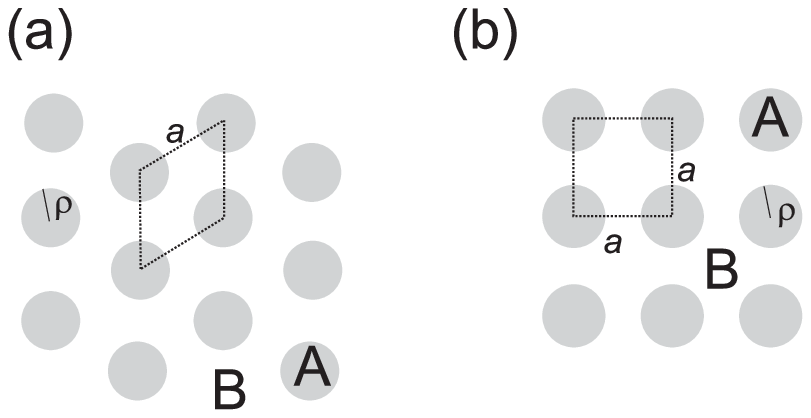}
\caption{\label{fig:1} Cross section of a 2D heterostructure consisting of infinitely long cylindrical rods (A) disposed in sites of a (a) triangular or (b) square lattice and embedded in matrix material (B). Dotted lines  represent unit cell limits, $a$ is the lattice constant, and $\rho$ the radius of the circular cross section of the rods; in both depicted structures the filling fraction is $f = 0.3$.}
\end{figure}

\begin{figure}
\centering
\includegraphics[width=3.5in]{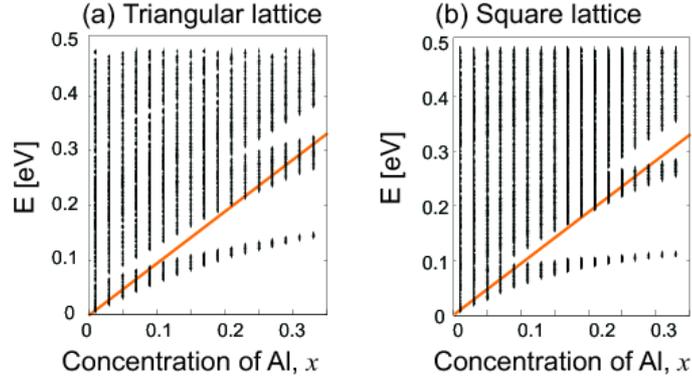}
\caption{\label{fig2x}Electronic band structure of 2D  superlattice formed by
periodic system of cylindrical GaAs rods disposed in sites of a triangular (a) or
square (b) lattice and embedded in Al$_{x}$Ga$_{1-x}$As, plotted versus matrix Al
concentration $x$ at fixed filling fraction $f = 0.3$ and lattice constant $a = 120$
{\AA}. The thick (red) line indicates the conduction band bottom energy in the
Al$_{x}$Ga$_{1-x}$As matrix.}
\end{figure}

\begin{figure}
\centering
\includegraphics[width=3.5in]{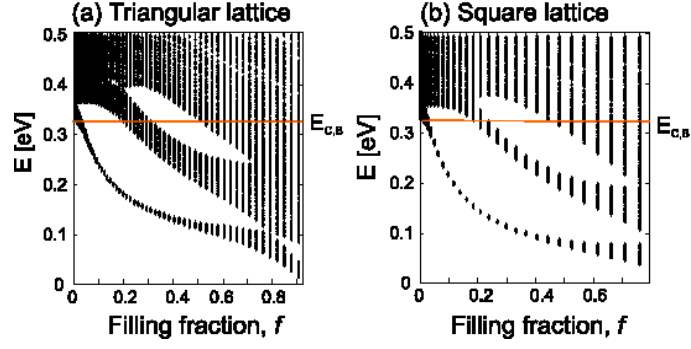}
\caption{\label{fig:3f}Electronic band structure  of 2D superlattice  formed by
periodic system of cylindrical GaAs rods disposed in sites of a triangular (a) or
square (b) lattice and embedded in Al$_{0.35}$Ga$_{0.65}$As, plotted versus filling
fraction $f$ at fixed lattice constant $a = 120$ {\AA}. The thick (red) horizontal
line indicates the conduction band bottom energy, $E_{C,B} \approx 0.33$, in the
matrix.}
\end{figure}

\begin{figure}
\centering
\includegraphics[width=9cm]{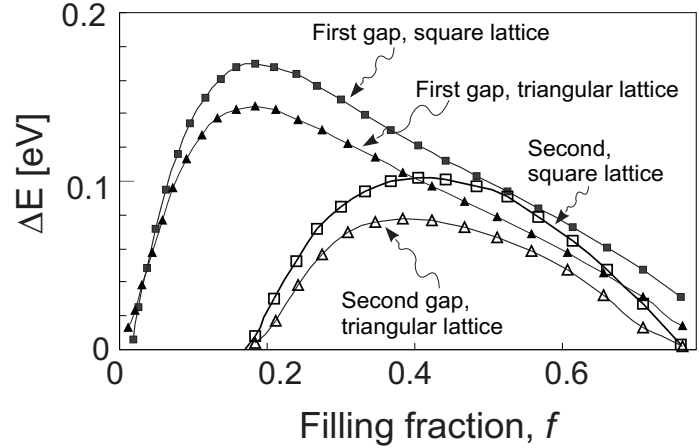}
\caption{\label{fig:4}
Width of first and  second energy minigap (solid and outlined symbols, respectively) versus
filling fraction $f$ in a triangular or square lattice-based heterostructure
(triangles and squares, respectively). Heterostructure form quantum wire arrays  of cylindrical GaAs rods disposed in sites of a triangular or square lattice and embedded in Al$_{0.35}$Ga$_{0.65}$As; lattice constant $a = 120$ {\AA}.}
\end{figure}

\begin{figure}[pth]
\includegraphics[width=3.5in]{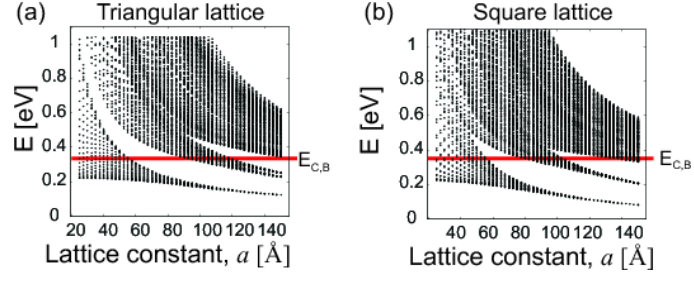}
\caption{\label{fig:5} Electronic  band structure of quantum wire arrays formed by
system of cylindrical GaAs rods disposed in sites of a triangular (a) or square (b)
lattice and embedded in Al$_{0.35}$Ga$_{0.65}$As, plotted versus lattice constant
$a$ at fixed filling fraction $f=0.3$. The thick (gray) horizontal line indicates the
conduction band bottom energy in the matrix material, $E_{C,B}$.}
\end{figure}

\begin{figure}[pth]
\includegraphics[width=2.5in]{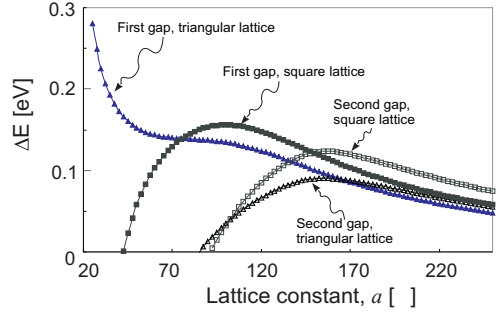}
\caption{\label{fig:6}Width of first  and second electronic minigap (filled and
outlined symbols, respectively) versus lattice constant $a$ in triangular or square
lattice-based heterostructure (triangles and squares, respectively) at fixed filling
fraction $f=0.3$. The heterostructure consists of a periodic system of cylindrical
GaAs rods embedded in Al$_{0.35}$Ga$_{0.65}$As.}
\end{figure}

\begin{figure}
\includegraphics[width=4.5in]{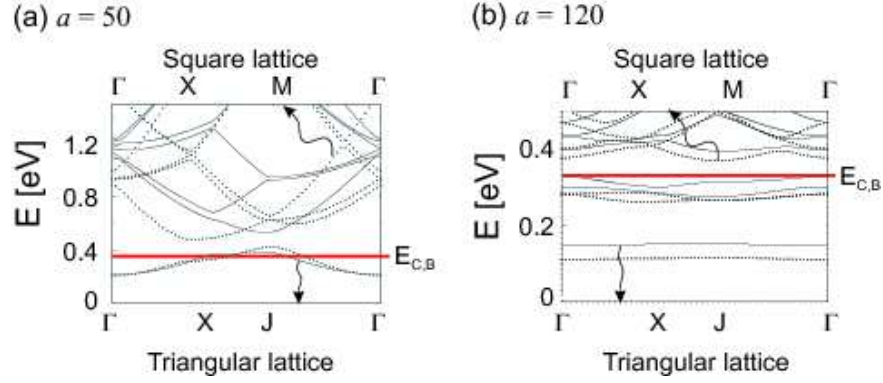}
\caption{\label{fig:7}Electronic minibands in system of GaAs rods disposed in sites
of triangular (solid lines) and square lattice (dashed lines) with lattice constant $a=50$ {\AA} (a) and $a= 120$ {\AA} (b).
Rods are embedded in Al$_{0.35}$Ga$_{0.65}$As, and the filling fraction is $f=0.3$.
Note the different energy scale in the two plots. The thick (gray) horizontal line
indicates the conduction band bottom energy in the matrix material, $E_{C,B}$.}
\end{figure}

\begin{figure}
\includegraphics[width=2.0in]{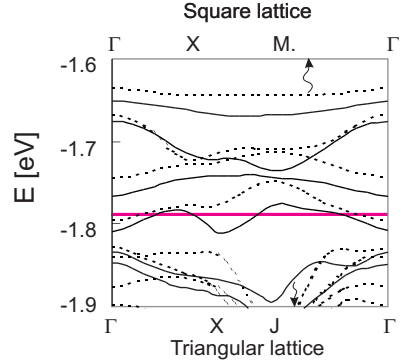}
\caption{\label{fig:8}Hole minibands in heterostructure with GaAs rods
disposed in sites of triangular or square lattice (solid and dashed lines,
respectively), calculated along the indicated path in the first Brillouin zone;
bottom and top scales refer to triangular and square lattice-based heterostructure,
respectively. In both cases the matrix material is Al$_{0.35}$Ga$_{0.65}$As, the
lattice constant $a = 50$ {\AA}, and the filling fraction $f=0.3$. The thick (red)
horizontal line indicates the valence band top energy, $E_{V,B} = -1.78$, in the
matrix material. }
\end{figure}

\begin{figure}
\includegraphics[width=3.5in]{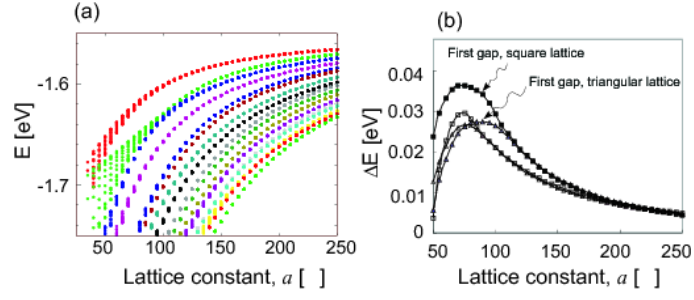}
\caption{\label{fig:9} (a) Hole minibands  versus lattice constant in
heterostructure with GaAs rods disposed in triangular lattice sites. (b) Width of
the first two hole minigaps, between minibands 1 and 2 (filled symbols) and between
minibands 3 and 4 (outlined symbols), versus $a$ for rods disposed in triangular or
square lattice sites (triangles and squares, respectively). The matrix material is
Al$_{0.35}$Ga$_{0.65}$As and the filling fraction $f=0.3$. The valence band top
energy is $E_{V,B} = -1.78$ in this case.  }
\end{figure}

\begin{figure}
\includegraphics[width=3.5in]{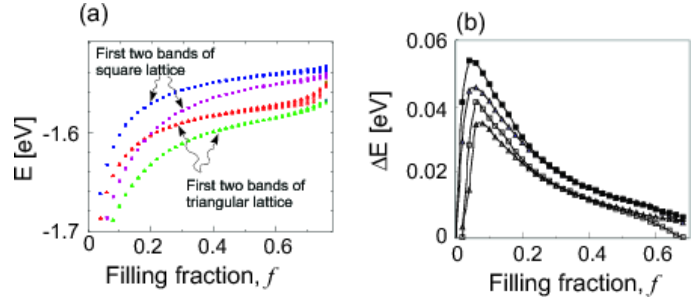}
\caption{\label{fig:10} (a) Energy of the first two hole minibands  versus filling
fraction in quantum wire array with GaAs rods disposed in triangular or square
lattice sites (triangles and squares, respectively). (b) Width of top two minigaps,
between minibands 1 and 2 (filled symbols) and between minibands 3 and 4 (outlined
symbols) versus $f$ in the triangular and square lattice case (triangles and
squares, respectively). The matrix material is Al$_{0.35}$Ga$_{0.65}$As, the lattice
constant $a = 50$ {\AA}, and the filling fraction $f=0.3$.}
\end{figure}

\end{document}